\documentclass[aps,prb,english,showpacs,twocolumn,superscriptaddress]{revtex4-2}
\usepackage{amsfonts}
\usepackage{amssymb}
\usepackage{amsmath}
\usepackage{graphicx}

\usepackage{epsfig}
\usepackage{subfigure}
\usepackage{color}
\usepackage{amsmath,bm}
\usepackage{booktabs}
\usepackage{appendix}
\usepackage[colorlinks=true, linkcolor=blue, citecolor=blue, urlcolor=blue]{hyperref}
\usepackage{orcidlink}
\begin{document}
	
	\title{String Order in Quench Dynamics as a Probe of Quantum Phase Transitions}
	
	\author{H. P. Zhang\,\orcidlink{0009-0001-5289-4865}}
	
	\affiliation{
		\href{https://gscaep.ac.cn/}
		{Graduate School of China Academy of Engineering Physics},
		Beijing 100193, China
	}
	
	\author{Z. Song\,\orcidlink{0000-0002-3315-4589}}
	\email[Corresponding author: ]{songtc@nankai.edu.cn}
	\affiliation{
		\href{https://www.nankai.edu.cn/}
		{School of Physics, Nankai University},
		Tianjin 300071, China
	}

\begin{abstract}
{We study the anisotropic XY chain in a transverse field by introducing
a global string order parameter and examining its equilibrium behavior and its quench dynamics from a fully polarized initial state.} The corresponding string operator is defined as the
product of local projected spin operators over the entire chain. {The nonequilibrium order parameter is the long-time average of the logarithm of the string-operator expectation value per site after the quench.} {For the ground state, its field derivative is singular on the Ising critical lines; after the quench, the steady-state response also records the field boundaries and the anisotropy transition.} {We derive analytically the nonanalytic behavior of the order parameter at the quantum critical lines.} Our results reveal
a hidden nonlocal order underlying this paradigmatic model of conventional
quantum phase transitions. {They also connect measurable spin populations with equilibrium and dynamical topological structure.} 
\end{abstract}
\maketitle

\section{Introduction}

\begingroup

Quantum phase transitions (QPTs) are changes of the ground state driven by quantum fluctuations at zero temperature. They organize phenomena ranging from magnetism and superconductivity to topological matter, and connect microscopic Hamiltonians to universal long-wavelength behavior \cite{Sachdev_1999,Stockert_2011,Breunig_2017,Shibauchi_2014,Kirschbaum_2026,Gross_2017}. One-dimensional spin chains are particularly useful because representative models are exactly solvable while retaining nontrivial critical behavior. The transverse-field Ising model and its anisotropic XY generalization are canonical examples in which phase boundaries are controlled by the closing of a quasiparticle gap \cite{Pfeuty_1970,Barouch_1970}.

Landau theory distinguishes phases through local order parameters and spontaneous symmetry breaking \cite{Dong_2017}, but this description does not capture all forms of quantum order. Symmetry-protected topological (SPT) phases and some nonequilibrium states can share the same local symmetries while differing in nonlocal correlations and entanglement structure \cite{Chen_2012,Senthil_2015,Gu_2009,Chen_2013,Kundu_2014,Dahlhaus_2015,Wilczek_2012,von_Keyserlingk_2016,Bakker_2022,Fazzini_2021,Berdanier_2018}. String order, introduced to expose hidden antiferromagnetism in the Haldane phase \cite{Haldane_1983,den_Nijs_1989}, has consequently become a standard diagnostic of one-dimensional SPT systems and has been related to matrix-product states, entanglement spectra, projective representations, and topological invariants \cite{Alcaraz_1992,P_rez_Garc_a_2008,Lou_2003,Kennedy_1992,Pollmann_2012,Haegeman_2012}.

The anisotropic XY chain provides a transparent setting for connecting such nonlocal observables to topology. Under the Jordan--Wigner transformation it maps to a spinless $p$-wave superconducting chain whose Bogoliubov Hamiltonian is characterized by a winding number; the phase boundaries occur when the quasiparticle gap closes \cite{Lieb_1961,Kitaev2001,Schnyder_2008,Barouch_1970,Asboth2016}. The fermionic representation exposes the independent momentum sectors, whereas the spin representation is natural for measurements. This mapping also clarifies why gap-closing modes can control response derivatives even when an extensive configuration probability remains smooth. A spin observable with this sensitivity would complement standard topological diagnostics. It is therefore useful to identify a simple spin observable whose singular response reveals this structure without reconstructing the winding number directly.

This question is especially relevant out of equilibrium. Ultracold atoms, trapped ions, superconducting qubits, and programmable quantum simulators now permit controlled studies of quenches, relaxation, prethermalization, and dynamical phase transitions \cite{Bloch_2008,Gross_2017,Sch_fer_2020,Altman_2021,Daley_2022}. Nonequilibrium transitions have been characterized through Loschmidt quantities, entanglement, topological indices, and correlation functions \cite{Heyl_2013,Jurcevic_2017,Fl_schner_2017,Calabrese_2005,Budich_2016,Andraschko_2014}, but nonlocal order in long-time states remains less explored. Conventional string correlators involve endpoint operators whose choice can complicate analytic dynamics and obscure their relation to ordinary critical lines.

Here we introduce the global string operator
\begin{equation*}
\hat O=\prod_{l=1}^{N}\frac{1+\sigma_l^z}{2},
\end{equation*}
which projects onto a fully polarized configuration. After the Jordan--Wigner transformation it becomes a product of fermion-occupation operators, so its expectation value in the states considered here factorizes over momentum sectors. We use its intensive logarithm as a string order parameter and evaluate it exactly for the ground state and for the long-time average after a quench from the fully polarized state. The global construction avoids endpoint choices and gives a direct configuration-space interpretation that is compatible with population measurements in quantum simulators. The ground-state field response develops logarithmic singularities on the Ising critical lines, while the post-quench order forms a field-independent plateau inside the gapped topological region, with singular field response at its boundaries and a finite anisotropy-derivative jump on the gapless line $\gamma=0$ for $|h|<1$.

This paper is organized as follows. Section II introduces the XY model, its fermionic momentum-sector decomposition, and the global string operator. Section III derives the ground-state string order and shows how its field derivative traces the critical lines. Section IV treats the quench dynamics and obtains the long-time-averaged order, including its singular field response, field-independent plateau in the gapped region, and anisotropy-derivative jump at the isotropic critical line. Section V summarizes the resulting connection between global string order, nonequilibrium dynamics, and the topological structure of the chain.
\endgroup

\section{Model and string operator}

\label{Model and phase diagram}

We consider a one-dimensional spin-$1/2$ XY model in a transverse
magnetic field $h$ on an $N$-site lattice. {We take $N$ to be even throughout.} The Hamiltonian is given
by

\begin{equation}
H=-\sum\limits^{N}_{j=1}\left(\frac{1+\gamma}{2}\sigma^{x}_{j}\sigma^{x}_{j+1}+\frac{1-\gamma}{2}\sigma^{y}_{j}\sigma^{y}_{j+1}\right)+\sum\limits^{N}_{j=1}h\sigma^{z}_{j},\label{H}
\end{equation}
where $\sigma^{\mu}_{j}$ ($\mu=x,y,z$) are the Pauli operators acting
on site $j$, and satisfy $\sigma^{\mu}_{j}=\sigma^{\mu}_{j+N}$ under
periodic boundary conditions, whereas the boundary couplings are removed
under open boundary conditions. Applying the Jordan-Wigner transformation
\cite{Jordan_1928},
\begin{eqnarray}
\sigma^{x}_{j} & = & \prod\limits^{j-1}_{i=1}\left(1-2c^{\dagger}_{i}c_{i}\right)\left(c^{\dagger}_{j}+c_{j}\right)\nonumber \\
\sigma^{y}_{j} & = & -i\prod\limits^{j-1}_{i=1}\left(1-2c^{\dagger}_{i}c_{i}\right)\left(c^{\dagger}_{j}-c_{j}\right)\nonumber \\
\sigma^{z}_{j} & = & 2c^{\dagger}_{j}c_{j}-1
\end{eqnarray}
the Pauli operators are mapped onto fermionic operators. The parity
of the number of fermions

\begin{equation}
\Pi=\prod^{N}_{l=1}\left(-\sigma^{z}_{l}\right)=\left(-1\right)^{N_{p}}
\end{equation}
is a conserved quantity, i.e., $\left[H,\Pi\right]=0$, where $N_{p}=\sum^{N}_{j=1}c^{\dag}_{j}c_{j}$.
Then the Hamiltonian (\ref{H}) can be rewritten as $H=\sum_{\eta=+,-}P_{\eta}H_{\eta}P_{\eta}$,
where $P_{\eta}=\frac{1}{2}\left(1+\eta\Pi\right)$ is the projector
onto the subspace with even ($\eta=+$) and odd ($\eta=-$) $N_{p}$.
Consequently, the Hamiltonian in each invariant subspace has the form

\begin{eqnarray}
H_{\eta} & = & -\sum\limits^{N-1}_{j=1}(c^{\dag}_{j}c_{j+1}+\gamma c^{\dag}_{j}c^{\dag}_{j+1})+\eta(c^{\dag}_{N}c_{1}+\gamma c^{\dag}_{N}c^{\dag}_{1})\nonumber \\
 &  & +\mathrm{H.c.}+2h\sum\limits^{N}_{j=1}c^{\dag}_{j}c_{j}-Nh.
\end{eqnarray}
The Hamiltonians $H_{\pm}$ can be written in block-diagonal form
by taking the Fourier transformation as $c_{j}=\frac{1}{\sqrt{N}}\sum_{k_{\pm}}e^{ik_{\pm}j}c_{k_{\pm}}$,
with the allowed momenta $k_{+}=2\left(m+1/2\right)\pi/N$, $k_{-}=2m\pi/N$,
{$m=-N/2,-N/2+1,\ldots,N/2-1$}. In the thermodynamic limit, the difference
between $k_{-}$ and $k_{+}$ can be ignored. In the following,
we neglect this distinction and use the same momentum set, denoted
by $k$, for both parity sectors. Under periodic boundary conditions,
the Hamiltonian can be decomposed into independent momentum sectors
as

\begin{equation}
H=\sum\limits_{\pi>k>0}H_{k},
\end{equation}
with
\begin{eqnarray}
H_{k} & = & 2\left(h-\cos k\right)c^{\dag}_{k}c_{k}+2\left(\cos k-h\right)c_{-k}c^{\dag}_{-k}\nonumber \\
 &  & -2i\gamma\sin k(c_{k}c_{-k}+c^{\dag}_{k}c^{\dag}_{-k}),\label{Kitaev model}
\end{eqnarray}
where the Hamiltonians corresponding to different momenta commute,
\begin{equation}
\left[H_{k},H_{k^{\prime}}\right]=0.
\end{equation}
Therefore, each momentum sector can be diagonalized independently,
and the corresponding dynamics can be treated separately.

\begin{figure*}
	\centering \includegraphics[width=\linewidth]{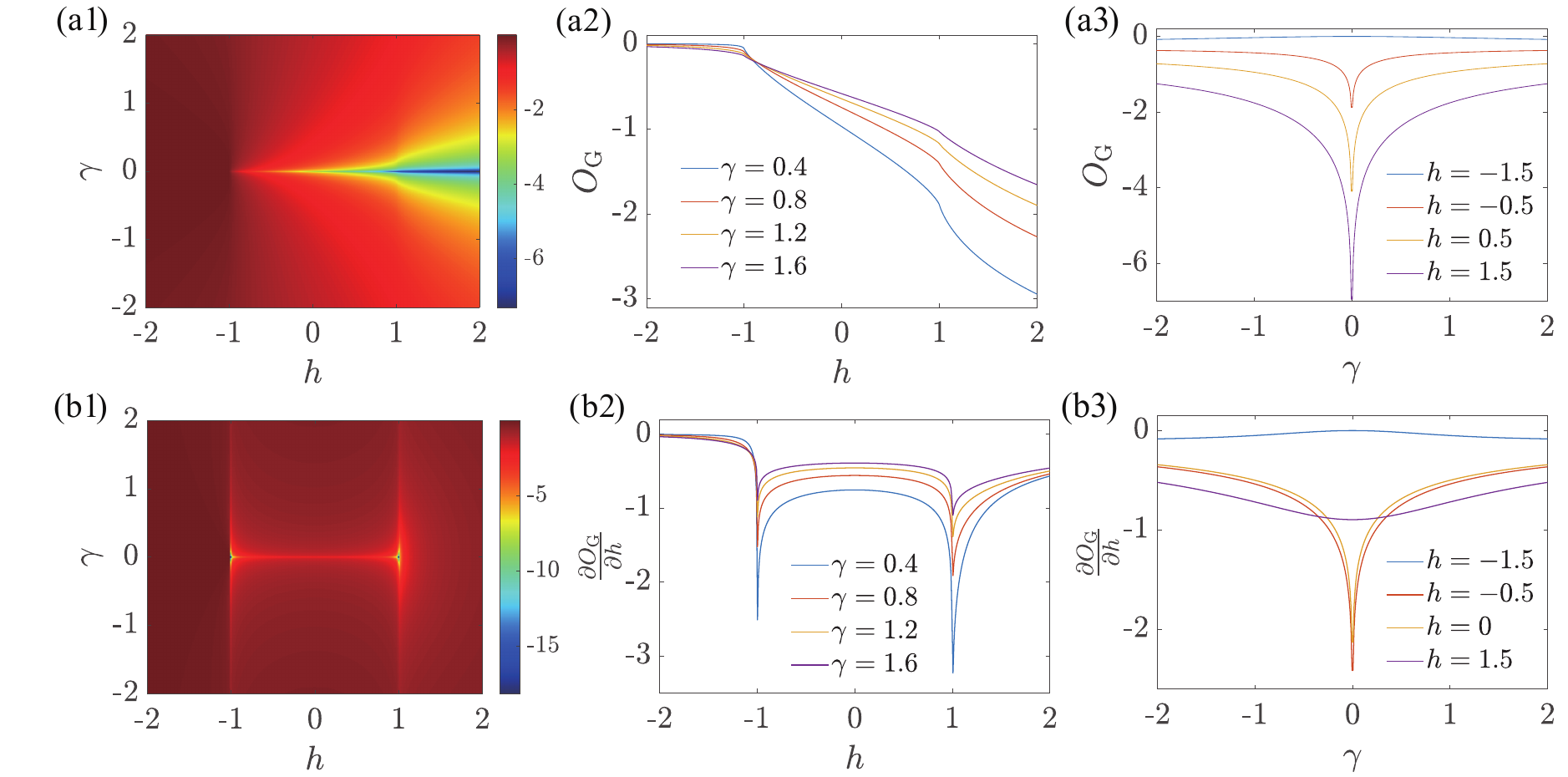}
	\caption{Ground-state intensive string order $O_{\mathrm G}$ and its transverse-field response in the anisotropic XY chain for $N=10\,000$. Panels (a1) and (b1) show color maps of $O_{\mathrm G}$ and $\partial O_{\mathrm G}/\partial h$, respectively, in the $(h,\gamma)$ plane. Panels (a2) and (b2) show their field dependence at fixed anisotropies, while panels (a3) and (b3) show their anisotropy dependence at the fields specified in the respective legends. The order is strongly suppressed near $\gamma=0$ for the displayed cuts with $h>-1$, whereas its field derivative develops pronounced negative extrema at $h=\pm1$ and near $\gamma=0$ for $|h|<1$. These features connect the parameter-space maps to the one-dimensional cuts and reveal the critical lines through the nonlocal response. The symmetry under $\gamma\to-\gamma$ contrasts with the field asymmetry of $O_{\mathrm G}$, which reflects the choice of the fully polarized reference configuration.}
	\label{fig:1}
\end{figure*}

{For a given momentum $k$, the Hamiltonian $H_k$ is represented in the basis $\{\lvert 0\rangle_k\lvert 0\rangle_{-k},\lvert 1\rangle_k\lvert 1\rangle_{-k},\lvert 1\rangle_k\lvert 0\rangle_{-k},\lvert 0\rangle_k\lvert 1\rangle_{-k}\}$ by the following $4\times4$ matrix:}
\begin{equation}
h_{k}=2\left(\begin{array}{cccc}
\cos k-h & i\gamma\sin k\\
-i\gamma\sin k & h-\cos k\\
 &  & 0 & 0\\
 &  & 0 & 0
\end{array}\right),
\end{equation}
where $\left\vert 1\right\rangle _{k}=c^{\dag}_{k}\left\vert 0\right\rangle _{k}$
and $c_{k}\left\vert 0\right\rangle _{k}=0$. {The corresponding eigenstates, where the labels $e$ and $o$ denote the even and odd parity sectors, respectively, are given by}

\begin{eqnarray}
\left\vert \varphi^{\pm}_{k\mathrm{e}}\right\rangle  & = & \frac{1}{\Omega^{\pm}}\left(a^{\pm}_{k}\left\vert 0\right\rangle _{k}\left\vert 0\right\rangle _{-k}+\left\vert 1\right\rangle _{k}\left\vert 1\right\rangle _{-k}\right),\nonumber \\
\left\vert \varphi^{+}_{k\mathrm{o}}\right\rangle  & = & \left\vert 1\right\rangle _{k}\left\vert 0\right\rangle _{-k},\left\vert \varphi^{-}_{k\mathrm{o}}\right\rangle =\left\vert 0\right\rangle _{k}\left\vert 1\right\rangle _{-k},\label{eigenstate}
\end{eqnarray}
where $\Omega^{\pm}=\sqrt{1+\left\vert a^{\pm}_{k}\right\vert ^{2}}$
is the normalization factor with
\begin{equation}
a^{\pm}_{k}=i\frac{\cos k-h+\epsilon^{\pm}_{k\mathrm{e}}/2}{\gamma\sin k}.
\end{equation}
The corresponding eigenenergies are 
\begin{equation}
\epsilon^{\pm}_{k\mathrm{e}}=\pm\varepsilon_{k},\epsilon^{\pm}_{k\mathrm{o}}=0,
\end{equation}
where
\begin{equation}
\varepsilon_{k}=2\sqrt{\left(h-\cos k\right)^{2}+\gamma^{2}\sin^{2}k}.
\end{equation}
Accordingly, the ground state is
\begin{equation}
\left\vert \text{\textrm{G}}\right\rangle =\prod_{\pi>k>0}\left\vert \varphi^{-}_{k\mathrm{e}}\right\rangle .\label{GS}
\end{equation}
{The phase boundaries are determined by the gap-closing condition $\min_k\varepsilon_k=0$, which yields the critical lines} 
\begin{equation}
h=\pm1\text{, and }\gamma=0\text{, }\left\vert h\right\vert <1.
\end{equation}
{These critical lines separate four gapped regions: the two regions with $|h|>1$ have winding number $0$, whereas those with $|h|<1$ and opposite signs of $\gamma$ have winding numbers $\pm1$.}

We define the string operator 
\begin{equation}
\hat{O}=\prod^{N}_{l=1}\frac{1+\sigma^{z}_{l}}{2},
\end{equation}
which is the product of local spin projection operators over the entire
chain. Unlike conventional string operators defined between two endpoints,
the present operator is a global string operator consisting of the
product of local projection operators over the entire chain. It projects
onto the fully polarized configuration and serves as a nonlocal observable
characterizing the many-body state. Since the expectation value of
the string operator decays exponentially with the system size, we
define the intensive string order parameter as 
\begin{equation}
O=\lim_{N\rightarrow\infty}\frac{1}{N}\ln\langle\psi|\hat{O}|\psi\rangle,\label{SOP}
\end{equation}
for a given state $|\psi\rangle$. {For even $N$, the fully polarized configuration has even fermion parity. Hence $\langle\psi|\hat O|\psi\rangle=0$ for any odd-parity state, and its logarithm is not finite. The following calculations use the even-parity sector.}

\section{Ground-state string order}

\label{Ground-state string order}

In this section, we investigate the properties of the ground state
by calculating the ground-state string order and its field response, which can identify the phase boundaries. The derivation is based on the fermionic
representation of the string operator,
\begin{equation}
\hat{O}=\prod^{N}_{l=1}c^{\dag}_{l}c_{l},
\end{equation}
obtained through the Jordan--Wigner transformation, which enables
an exact analytical evaluation of the string order parameter. Below,
we present the analytical derivation for the behavior of the string
order in the $\left(\gamma,h\right)$ parameter plane.

{Based on the ground-state wave function given in Eq.~(\ref{GS}),}
the string order parameter in the thermodynamic limit is

\begin{equation}
O_{\text{\textrm{G}}}=\lim_{N\rightarrow\infty}\frac{1}{N}\ln\left\langle \text{\textrm{G}}\right\vert \hat{O}\left\vert \text{\textrm{G}}\right\rangle =\frac{1}{2\pi}\int^{\pi}_{0}\ln O_{k}\mathrm{d}k
\end{equation}
where
\begin{equation}
O_{k}=\frac{\gamma^{2}\sin^{2}k}{\varepsilon_{k}\left[\varepsilon_{k}/2-\left(\cos k-h\right)\right]}.\label{ground_sector_weight}
\end{equation}
A direct calculation yields
\begin{equation}
\frac{\partial O_{\text{\textrm{G}}}}{\partial h}=-\frac{1}{\pi}\int^{\pi}_{0}\frac{\mathrm{d}k}{\varepsilon_{k}}+F(\gamma,h)
\end{equation}
where
\begin{equation}
F(\gamma,h)=\frac{1}{2\pi}\int^{\pi}_{0}\frac{\cos k-h}{\left(\cos k-h\right)^{2}+\gamma^{2}\sin^{2}k}\mathrm{d}k.
\end{equation}
\begingroup

The first integral can be expressed in terms of complete elliptic integrals and exhibits logarithmic divergences as $h\to\pm1$ at fixed $\gamma\ne0$, and as $\gamma\to0$ for $|h|<1$. For nonzero $\gamma$, the residue theorem gives a function $F(\gamma,h)$ that is analytic away from $h=\pm1$, where it has finite jumps. In finite chains, the divergent field response is resolved as sharp negative extrema near the critical lines.

To assess which features of the nonlocal observable carry the critical information, we evaluate the order and its field response throughout the $(h,\gamma)$ plane. The resulting maps and representative cuts are plotted in Fig.~\ref{fig:1}. The order is even under $\gamma\to-\gamma$ but asymmetric in $h$, reflecting the choice of a fixed polarized projector rather than a change of phase. By contrast, its field derivative develops narrow negative structures at the Ising boundaries and becomes increasingly pronounced on approaching $\gamma=0$ from within $|h|<1$, while the order itself remains smooth at the boundaries for nonzero $\gamma$. Outside this interval, the field response remains regular near the isotropic limit even though the projection weight can be strongly suppressed. The locations of the finite-size extrema track the logarithmic singularities of the momentum integral and provide a numerical signature of the gap-closing boundaries. Thus the field response separates the critical nonanalyticity from the state-selection bias inherent in the global projector.
\endgroup

\section{Steady-state string order}

\label{Steady-state string order}

\begin{figure*}
	\centering \includegraphics[width=0.95\linewidth]{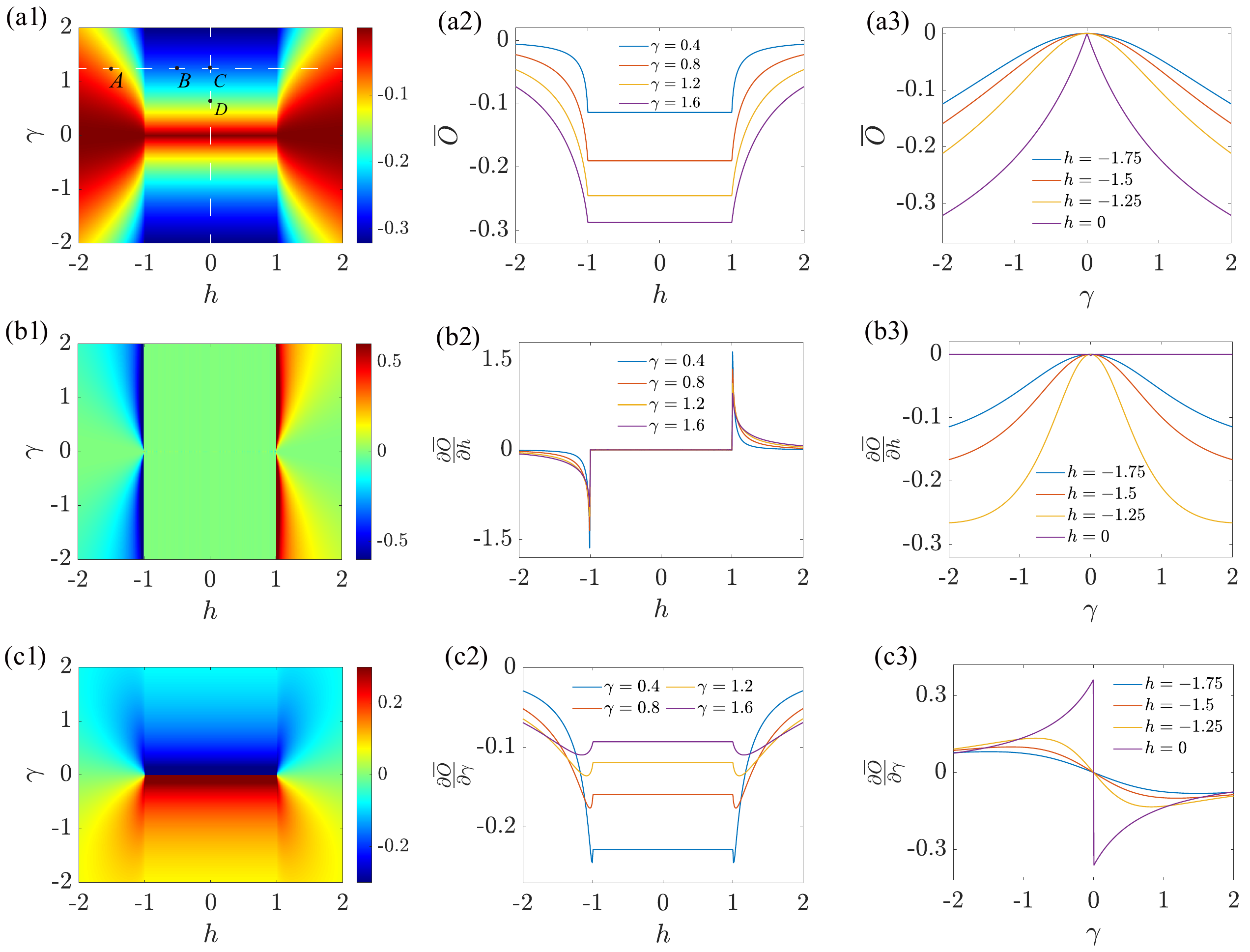}
	\caption{Nonequilibrium intensive string order after a quench from the fully polarized state, for $N=10\,000$ and a time-averaging window $T=1000$. Panels (a1)--(a3) show the averaged order $\overline O$: (a1) is the $(h,\gamma)$ map, and (a2) and (a3) give its field and anisotropy dependence, respectively, with fixed parameters specified in the legends. Panels (b1)--(b3) present the field derivative $\partial\overline O/\partial h$ in the same three representations, and panels (c1)--(c3) present the anisotropy derivative $\partial\overline O/\partial\gamma$. The dashed lines at $\gamma=1.2$ and $h=0$ in (a1) specify the scans in Fig.~\ref{fig:3}(a,b); points $A$--$D$ select the running averages in Fig.~\ref{fig:3}(c). The field-independent plateau for $|h|<1$ is bounded by an enhanced field response at $h=\pm1$, while the anisotropy response has a jump across $\gamma=0$ within $|h|<1$. Together, these features distinguish the Ising boundaries from the gapless isotropic line across which the winding orientation reverses.}
	\label{fig:2}
\end{figure*}

In this section, we study the long-time nonequilibrium dynamics of
the string order parameter following a quantum quench from the fully
polarized state. Specifically, we consider the time evolution $\left\vert \psi(t)\right\rangle =e^{-iHt}\left\vert \psi(0)\right\rangle $
starting from the fully polarized initial state

\begin{equation}
\left\vert \psi(0)\right\rangle =\prod^{N}_{j=1}\left\vert \uparrow\right\rangle _{j}.
\end{equation}
The long-time-averaged string order parameter in the thermodynamic
limit is defined as
\begin{align}
\overline{O} & =\lim_{T,N\rightarrow\infty}\frac{1}{T}\int^{T}_{0}O(t)\text{\textrm{d}}t \notag \\
 & =\lim_{T,N\rightarrow\infty}\frac{1}{T}\int^{T}_{0}\frac{1}{N}\ln\left\langle \psi(t)\right\vert \hat{O}\left\vert \psi(t)\right\rangle \text{\textrm{d}}t,
 \label{O_bar}
\end{align}
which can be rewritten as

\begin{equation}
\overline{O}=\lim_{T\rightarrow\infty}\frac{1}{2\pi T}\int^{\pi}_{0}\mathrm{d}k\int^{T}_{0}\ln O_{k}(t)\mathrm{d}t.
\label{Obar}
\end{equation}
Using the eigenstates of $H_{k}$, given in Eq. (\ref{eigenstate}),
we obtain the time-dependent string order parameter
\begin{equation}
O_{k}(t)=1-\sin^{2}\theta_{k}\sin^{2}\left(\varepsilon_{k}t\right),\label{sector_dynamics}
\end{equation}
where 
\begin{equation}
\theta_{k}=\tan^{-1}\frac{\gamma\sin k}{\cos k-h}.
\end{equation}
Due to the periodicity of $O_{k}(t)$, the long-time average can be
evaluated analytically as
\begin{equation}
\overline{O}=\frac{1}{\pi}\int^{\pi}_{0}\ln\frac{1+\left\vert \cos\theta_{k}\right\vert}{2}\,\mathrm{d}k.\label{averaged_order}
\end{equation}
Differentiating $\overline{O}$ with respect to the transverse field
$h$ and the anisotropy parameter $\gamma$, we obtain

\begin{equation}
\frac{\partial\overline{O}}{\partial h}=\frac{8\gamma^{2}}{\pi}\int^{\pi}_{0}\frac{\text{\textrm{sgn}}\left(h-\cos k\right)\sin^{2}k}{\left(1+\left\vert \cos\theta_{k}\right\vert \right)\varepsilon^{3}_{k}}\mathrm{d}k,
\end{equation}
and
\begin{equation}
\frac{\partial\overline{O}}{\partial\gamma}=-\frac{8\gamma}{\pi}\int^{\pi}_{0}\frac{\left\vert h-\cos k\right\vert \sin^{2}k}{\left(1+\left\vert \cos\theta_{k}\right\vert \right)\varepsilon^{3}_{k}}\mathrm{d}k.
\end{equation}
\begingroup

At fixed $\gamma\ne0$, the field response contains the kernel $\sin^2 k/[(1+|\cos\theta_k|)\varepsilon_k^3]$, which behaves as $k^{-1}$ at $h=1$ and as $(\pi-k)^{-1}$ at $h=-1$. This gives a logarithmically divergent field response upon approaching $|h|=1$ from $|h|>1$. Inside $|h|<1$, evaluation of the integral gives $\partial\overline O/\partial h=0$: the averaged order forms a field-independent plateau. This is consistent with the invariance of Eq.~(\ref{averaged_order}) under $h\to-h$ with $k\to\pi-k$. The anisotropy derivative is odd in $\gamma$ and has finite, nonzero one-sided limits at $\gamma=0$ for $|h|<1$, producing a jump across the gapless isotropic line where the winding orientation reverses.

To confront the long-time expression with finite-size numerical data, the averaged order and its two parameter derivatives are plotted in Fig.~\ref{fig:2}. A broad plateau develops throughout the interior $|h|<1$, whereas the order decreases outside this region and its field response is enhanced on approaching the Ising boundaries from the exterior. The anisotropy dependence carries a different signature: inside the topological region the order has a cusp at $\gamma=0$, with finite one-sided derivatives, while the corresponding exterior behavior is smooth. The finite averaging window rounds the singular structures without shifting their locations. These trends reproduce the analytical result: the field derivative vanishes identically on the interior plateau, acquires a logarithmic boundary enhancement at $|h|=1$, and the anisotropy derivative jumps across the gapless isotropic line. The two responses therefore distinguish the Ising boundaries from the change of winding orientation along $\gamma=0$.
\endgroup

\begin{figure*}
\centering \includegraphics[width=\linewidth]{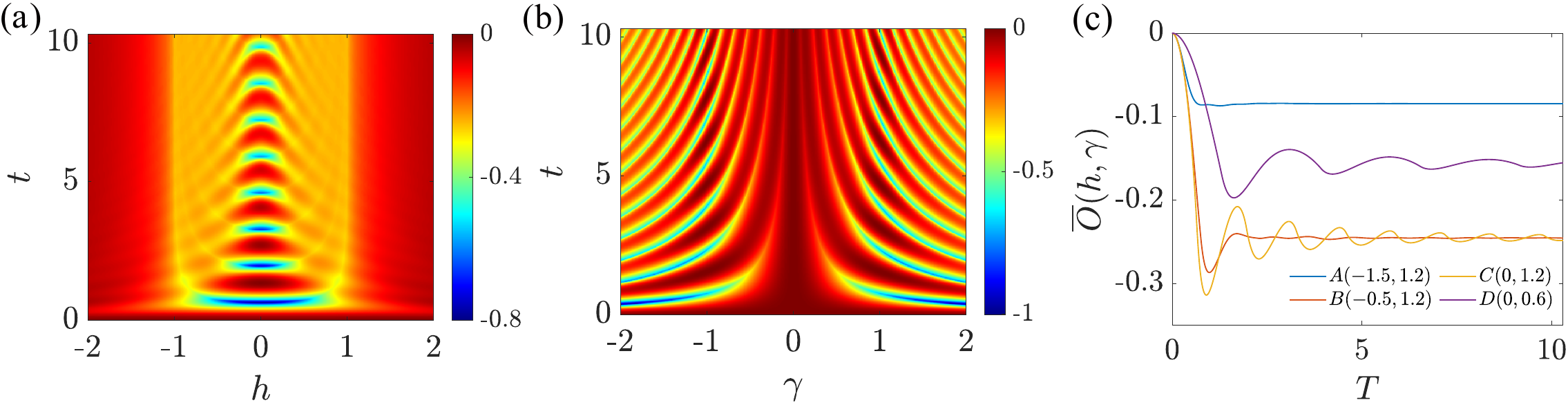}
\caption{Time evolution of the intensive string order $O(t)$ after a quench from the fully polarized state, and the associated running time average $\overline O(T)=T^{-1}\int_0^T O(t)\,\mathrm{d}t$, for a chain of $N=10\,000$ sites. Panels (a) and (b) show density plots of $O(t)$ as the transverse field $h$ and anisotropy $\gamma$ are varied, at fixed $\gamma=1.2$ and $h=0$, respectively. The pronounced time-dependent bands show that the transient response depends strongly on the post-quench parameters; near $\gamma=0$, the order is nearly time independent. Panel (c) compares the running averages at the four points $A$--$D$ marked in Fig.~\ref{fig:2}(a1). Their initial relaxation and subsequent residual oscillations differ markedly, illustrating that the route toward the long-time value depends on the quench point even when the steady-state order is well defined. The long-time averages connect the transient dynamics in this figure to the parameter-space results of Fig.~\ref{fig:2}.}
\label{fig:3}
\end{figure*}

\begingroup

The long-time average alone does not show how coherent oscillations are averaged out, or how rapidly a finite-time calculation approaches its asymptotic value. To make this distinction explicit, representative time evolutions together with their running averages are plotted in Fig.~\ref{fig:3}. The transients remain oscillatory and their amplitude and characteristic time depend on the quench point: fields far from the central region suppress the modulation, whereas reducing $|\gamma|$ changes the visible time scale and eventually suppresses the modulation near the isotropic line. At $\gamma=0$, the initial polarized state is an eigenstate of the number-conserving Hamiltonian, so the string order remains zero. Time averaging removes most of the oscillatory component, and quenches with the same anisotropy inside $|h|<1$ approach the same asymptotic value even when their transient traces differ; changing $\gamma$ changes that value. This time-domain convergence agrees with the field-independent interior of Eq.~(\ref{averaged_order}) and explains why the parameter-space plateau survives the coherent dynamics without requiring the instantaneous order to become stationary.
\endgroup

\section{Summary}

\label{sec_summary}

\begingroup

We introduced a global string operator in the anisotropic XY chain and its intensive logarithmic expectation value. The Jordan--Wigner transformation makes this quantity exactly tractable in ground states and in the long-time average after a quench. In equilibrium, its transverse-field derivative has logarithmic singularities on the Ising critical lines, while finite-size extrema track the phase boundaries. After a quench from the fully polarized state, the field response is singular on approaching the Ising lines from $|h|>1$ and vanishes inside the gapped topological region, where the averaged order is independent of $h$. The anisotropy derivative has a finite jump across the gapless line $\gamma=0$ for $|h|<1$, reflecting reversal of the Bogoliubov winding orientation. These results establish global string order as a compact nonlocal probe of gap closing and topological structure. Defined directly in spin space, it can also be tested through global population measurements in programmable quantum simulators and extended to other integrable chains.
\endgroup

\section*{Acknowledgment}

This work was supported by the National Natural Science Foundation of China
(under Grant No.~12374461).

\section*{DATA AVAILABILITY}

The data that support the findings of this article are not publicly
available. The data are available from the authors upon reasonable request.

\bibliography{StringOperatorReference}

@Article{Sachdev_1999,
  author    = {Sachdev, Subir},
  journal   = {Physics World},
  title     = {Quantum phase transitions},
  year      = {1999},
  issn      = {2058-7058},
  month     = Apr,
  number    = {4},
  pages     = {33--38},
  volume    = {12},
  doi       = {10.1088/2058-7058/12/4/23},
  publisher = {IOP Publishing},
}

@Article{Lieb_1961,
  author    = {Lieb, Elliott and Schultz, Theodore and Mattis, Daniel},
  journal   = {Annals of Physics},
  title     = {Two soluble models of an antiferromagnetic chain},
  year      = {1961},
  issn      = {0003-4916},
  month     = Dec,
  number    = {3},
  pages     = {407--466},
  volume    = {16},
  doi       = {10.1016/0003-4916(61)90115-4},
  publisher = {Elsevier BV},
}

@Article{Jordan_1928,
  author    = {Jordan, P. and Wigner, E.},
  journal   = {Zeitschrift fur Physik},
  title     = {Uber das Paulische Aquivalenzverbot},
  year      = {1928},
  issn      = {1434-601X},
  month     = Sep,
  number    = {9-10},
  pages     = {631--651},
  volume    = {47},
  doi       = {10.1007/bf01331938},
  publisher = {Springer Science and Business Media LLC},
}

@Article{Kitaev2001,
  author    = {Kitaev, A Yu},
  journal   = {Physics-Uspekhi},
  title     = {Unpaired Majorana fermions in quantum wires},
  year      = {2001},
  issn      = {1468-4780},
  month     = oct,
  number    = {10S},
  pages     = {131--136},
  volume    = {44},
  doi       = {10.1070/1063-7869/44/10s/s29},
  publisher = {Uspekhi Fizicheskikh Nauk (UFN) Journal},
}

@Article{Schnyder_2008,
  author    = {Schnyder, Andreas P. and Ryu, Shinsei and Furusaki, Akira and Ludwig, Andreas W. W.},
  journal   = {Physical Review B},
  title     = {Classification of topological insulators and superconductors in three spatial dimensions},
  year      = {2008},
  issn      = {1550-235X},
  month     = Nov,
  number    = {19},
  pages     = {195125},
  volume    = {78},
  doi       = {10.1103/physrevb.78.195125},
  publisher = {American Physical Society (APS)},
}

@Article{Haldane_1983,
  author    = {Haldane, F.D.M.},
  journal   = {Physics Letters A},
  title     = {Continuum dynamics of the 1-D Heisenberg antiferromagnet: Identification with the O(3) nonlinear sigma model},
  year      = {1983},
  issn      = {0375-9601},
  month     = Feb,
  number    = {9},
  pages     = {464--468},
  volume    = {93},
  doi       = {10.1016/0375-9601(83)90631-x},
  publisher = {Elsevier BV},
}

@Article{den_Nijs_1989,
  author    = {den Nijs, Marcel and Rommelse, Koos},
  journal   = {Physical Review B},
  title     = {Preroughening transitions in crystal surfaces and valence-bond phases in quantum spin chains},
  year      = {1989},
  issn      = {0163-1829},
  month     = Sep,
  number    = {7},
  pages     = {4709--4734},
  volume    = {40},
  doi       = {10.1103/physrevb.40.4709},
  publisher = {American Physical Society (APS)},
}

@Article{Shibauchi_2014,
  author    = {Shibauchi, T. and Carrington, A. and Matsuda, Y.},
  journal   = {Annual Review of Condensed Matter Physics},
  title     = {A Quantum Critical Point Lying Beneath the Superconducting Dome in Iron Pnictides},
  year      = {2014},
  issn      = {1947-5462},
  month     = Mar,
  number    = {1},
  pages     = {113--135},
  volume    = {5},
  doi       = {10.1146/annurev-conmatphys-031113-133921},
  publisher = {Annual Reviews},
}

@Article{Stockert_2011,
  author    = {Stockert, O. and Steglich, F.},
  journal   = {Annual Review of Condensed Matter Physics},
  title     = {Unconventional Quantum Criticality in Heavy-Fermion Compounds},
  year      = {2011},
  issn      = {1947-5462},
  month     = Mar,
  number    = {1},
  pages     = {79--99},
  volume    = {2},
  doi       = {10.1146/annurev-conmatphys-062910-140546},
  publisher = {Annual Reviews},
}

@Article{Breunig_2017,
  author    = {Breunig, Oliver and Garst, Markus and Klümper, Andreas and Rohrkamp, Jens and Turnbull, Mark M. and Lorenz, Thomas},
  journal   = {Science Advances},
  title     = {Quantum criticality in the spin-1/2 Heisenberg chain system copper pyrazine dinitrate},
  year      = {2017},
  issn      = {2375-2548},
  month     = Dec,
  number    = {12},
  volume    = {3},
  doi       = {10.1126/sciadv.aao3773},
  publisher = {American Association for the Advancement of Science (AAAS)},
}

@Article{Kirschbaum_2026,
  author    = {Kirschbaum, D. M. and Chen, L. and Zocco, D. A. and Hu, H. and Mazza, F. and Karlich, M. and Lužnik, M. and Nguyen, D. H. and Larrea Jiménez, J. and Strydom, A. M. and Adroja, D. and Yan, X. and Prokofiev, A. and Si, Q. and Paschen, S.},
  journal   = {Nature Physics},
  title     = {Emergent topological semimetal from quantum criticality},
  year      = {2026},
  issn      = {1745-2481},
  month     = Jan,
  number    = {2},
  pages     = {218--224},
  volume    = {22},
  doi       = {10.1038/s41567-025-03135-w},
  publisher = {Springer Science and Business Media LLC},
}

@Article{Pfeuty_1970,
  author    = {Pfeuty, Pierre},
  journal   = {Annals of Physics},
  title     = {The one-dimensional Ising model with a transverse field},
  year      = {1970},
  issn      = {0003-4916},
  month     = Mar,
  number    = {1},
  pages     = {79--90},
  volume    = {57},
  doi       = {10.1016/0003-4916(70)90270-8},
  publisher = {Elsevier BV},
}

@Article{Barouch_1970,
  author    = {Barouch, Eytan and McCoy, Barry M. and Dresden, Max},
  journal   = {Physical Review A},
  title     = {Statistical Mechanics of the XY Model. I},
  year      = {1970},
  issn      = {0556-2791},
  month     = Sep,
  number    = {3},
  pages     = {1075--1092},
  volume    = {2},
  doi       = {10.1103/physreva.2.1075},
  publisher = {American Physical Society (APS)},
}

@Article{Dong_2017,
  author    = {Dong Guo-Hui, Fang Yi-Nan and Sun Chang-Pu},
  journal   = {Communications in Theoretical Physics},
  title     = {Quantifying Spontaneously Symmetry Breaking of Quantum Many-Body Systems*},
  year      = {2017},
  issn      = {1572-9494},
  month     = Oct,
  number    = {4},
  pages     = {405},
  volume    = {68},
  doi       = {10.1088/0253-6102/68/4/405},
  publisher = {IOP Publishing},
}

@Article{Chen_2012,
  author    = {Chen, Xie and Gu, Zheng-Cheng and Liu, Zheng-Xin and Wen, Xiao-Gang},
  journal   = {Science},
  title     = {Symmetry-Protected Topological Orders in Interacting Bosonic Systems},
  year      = {2012},
  issn      = {1095-9203},
  month     = Dec,
  number    = {6114},
  pages     = {1604--1606},
  volume    = {338},
  doi       = {10.1126/science.1227224},
  publisher = {American Association for the Advancement of Science (AAAS)},
}

@Article{Senthil_2015,
  author    = {Senthil, T.},
  journal   = {Annual Review of Condensed Matter Physics},
  title     = {Symmetry-Protected Topological Phases of Quantum Matter},
  year      = {2015},
  issn      = {1947-5462},
  month     = Mar,
  number    = {1},
  pages     = {299--324},
  volume    = {6},
  doi       = {10.1146/annurev-conmatphys-031214-014740},
  publisher = {Annual Reviews},
}

@Article{Gu_2009,
  author    = {Gu, Zheng-Cheng and Wen, Xiao-Gang},
  journal   = {Physical Review B},
  title     = {Tensor-entanglement-filtering renormalization approach and symmetry-protected topological order},
  year      = {2009},
  issn      = {1550-235X},
  month     = Oct,
  number    = {15},
  pages     = {155131},
  volume    = {80},
  doi       = {10.1103/physrevb.80.155131},
  publisher = {American Physical Society (APS)},
}

@Article{Chen_2013,
  author    = {Chen, Xie and Gu, Zheng-Cheng and Liu, Zheng-Xin and Wen, Xiao-Gang},
  journal   = {Physical Review B},
  title     = {Symmetry protected topological orders and the group cohomology of their symmetry group},
  year      = {2013},
  issn      = {1550-235X},
  month     = Apr,
  number    = {15},
  pages     = {155114},
  volume    = {87},
  doi       = {10.1103/physrevb.87.155114},
  publisher = {American Physical Society (APS)},
}

@Article{Kundu_2014,
  author    = {Kundu, Arijit and Fertig, H. A. and Seradjeh, Babak},
  journal   = {Physical Review Letters},
  title     = {Effective Theory of Floquet Topological Transitions},
  year      = {2014},
  issn      = {1079-7114},
  month     = Dec,
  number    = {23},
  pages     = {236803},
  volume    = {113},
  doi       = {10.1103/physrevlett.113.236803},
  publisher = {American Physical Society (APS)},
}

@Article{Dahlhaus_2015,
  author    = {Dahlhaus, Jan P. and Fregoso, Benjamin M. and Moore, Joel E.},
  journal   = {Physical Review Letters},
  title     = {Magnetization Signatures of Light-Induced Quantum Hall Edge States},
  year      = {2015},
  issn      = {1079-7114},
  month     = Jun,
  number    = {24},
  pages     = {246802},
  volume    = {114},
  doi       = {10.1103/physrevlett.114.246802},
  publisher = {American Physical Society (APS)},
}

@Article{Wilczek_2012,
  author    = {Wilczek, Frank},
  journal   = {Physical Review Letters},
  title     = {Quantum Time Crystals},
  year      = {2012},
  issn      = {1079-7114},
  month     = Oct,
  number    = {16},
  pages     = {160401},
  volume    = {109},
  doi       = {10.1103/physrevlett.109.160401},
  publisher = {American Physical Society (APS)},
}

@Article{von_Keyserlingk_2016,
  author    = {von Keyserlingk, C. W. and Sondhi, S. L.},
  journal   = {Physical Review B},
  title     = {Phase structure of one-dimensional interacting Floquet systems. I. Abelian symmetry-protected topological phases},
  year      = {2016},
  issn      = {2469-9969},
  month     = Jun,
  number    = {24},
  pages     = {245145},
  volume    = {93},
  doi       = {10.1103/physrevb.93.245145},
  publisher = {American Physical Society (APS)},
}

@Article{Bakker_2022,
  author    = {Bakker, L. R. and Bahovadinov, M. S. and Kurlov, D. V. and Gritsev, V. and Fedorov, A. K. and Krimer, Dmitry O.},
  journal   = {Physical Review Letters},
  title     = {Driven-Dissipative Time Crystalline Phases in a Two-Mode Bosonic System with Kerr Nonlinearity},
  year      = {2022},
  issn      = {1079-7114},
  month     = Dec,
  number    = {25},
  pages     = {250401},
  volume    = {129},
  doi       = {10.1103/physrevlett.129.250401},
  publisher = {American Physical Society (APS)},
}

@Article{Fazzini_2021,
  author    = {Fazzini, Serena and Chudzinski, Piotr and Dauer, Christoph and Schneider, Imke and Eggert, Sebastian},
  journal   = {Physical Review Letters},
  title     = {Nonequilibrium Floquet Steady States of Time-Periodic Driven Luttinger Liquids},
  year      = {2021},
  issn      = {1079-7114},
  month     = Jun,
  number    = {24},
  pages     = {243401},
  volume    = {126},
  doi       = {10.1103/physrevlett.126.243401},
  publisher = {American Physical Society (APS)},
}

@Article{Berdanier_2018,
  author    = {Berdanier, William and Kolodrubetz, Michael and Parameswaran, S. A. and Vasseur, Romain},
  journal   = {Proceedings of the National Academy of Sciences},
  title     = {Floquet quantum criticality},
  year      = {2018},
  issn      = {1091-6490},
  month     = Aug,
  number    = {38},
  pages     = {9491--9496},
  volume    = {115},
  doi       = {10.1073/pnas.1805796115},
  publisher = {National Academy of Sciences},
}

@Article{Alcaraz_1992,
  author    = {Alcaraz, F. C. and Hatsugai, Y.},
  journal   = {Physical Review B},
  title     = {String correlation functions in the anisotropic spin-1 Heisenberg chain},
  year      = {1992},
  issn      = {1095-3795},
  month     = Dec,
  number    = {21},
  pages     = {13914--13918},
  volume    = {46},
  doi       = {10.1103/physrevb.46.13914},
  publisher = {American Physical Society (APS)},
}

@Article{P_rez_Garc_a_2008,
  author    = {Pérez-García, D. and Wolf, M. M. and Sanz, M. and Verstraete, F. and Cirac, J. I.},
  journal   = {Physical Review Letters},
  title     = {String Order and Symmetries in Quantum Spin Lattices},
  year      = {2008},
  issn      = {1079-7114},
  month     = Apr,
  number    = {16},
  pages     = {167202},
  volume    = {100},
  doi       = {10.1103/physrevlett.100.167202},
  publisher = {American Physical Society (APS)},
}

@Article{Lou_2003,
  author    = {Lou, Jizhong and Qin, Shaojin and Chen, Changfeng},
  journal   = {Physical Review Letters},
  title     = {String Order in Half-Integer-Spin Antiferromagnetic Heisenberg Chains},
  year      = {2003},
  issn      = {1079-7114},
  month     = Aug,
  number    = {8},
  pages     = {087204},
  volume    = {91},
  doi       = {10.1103/physrevlett.91.087204},
  publisher = {American Physical Society (APS)},
}

@Article{Kennedy_1992,
  author    = {Kennedy, Tom and Tasaki, Hal},
  journal   = {Physical Review B},
  title     = {Hidden ${Z}_2\times {Z}_2$ symmetry breaking in Haldane-gap antiferromagnets},
  year      = {1992},
  issn      = {1095-3795},
  month     = Jan,
  number    = {1},
  pages     = {304--307},
  volume    = {45},
  doi       = {10.1103/physrevb.45.304},
  publisher = {American Physical Society (APS)},
}

@Article{Pollmann_2012,
  author    = {Pollmann, Frank and Turner, Ari M.},
  journal   = {Physical Review B},
  title     = {Detection of symmetry-protected topological phases in one dimension},
  year      = {2012},
  issn      = {1550-235X},
  month     = Sep,
  number    = {12},
  pages     = {125441},
  volume    = {86},
  doi       = {10.1103/physrevb.86.125441},
  publisher = {American Physical Society (APS)},
}

@Article{Haegeman_2012,
  author    = {Haegeman, Jutho and Pérez-García, David and Cirac, Ignacio and Schuch, Norbert},
  journal   = {Physical Review Letters},
  title     = {Order Parameter for Symmetry-Protected Phases in One Dimension},
  year      = {2012},
  issn      = {1079-7114},
  month     = Jul,
  number    = {5},
  pages     = {050402},
  volume    = {109},
  doi       = {10.1103/physrevlett.109.050402},
  publisher = {American Physical Society (APS)},
}

@Article{Bloch_2008,
  author    = {Bloch, Immanuel and Dalibard, Jean and Zwerger, Wilhelm},
  journal   = {Reviews of Modern Physics},
  title     = {Many-body physics with ultracold gases},
  year      = {2008},
  issn      = {1539-0756},
  month     = Jul,
  number    = {3},
  pages     = {885--964},
  volume    = {80},
  doi       = {10.1103/revmodphys.80.885},
  publisher = {American Physical Society (APS)},
}

@Article{Gross_2017,
  author    = {Gross, Christian and Bloch, Immanuel},
  journal   = {Science},
  title     = {Quantum simulations with ultracold atoms in optical lattices},
  year      = {2017},
  issn      = {1095-9203},
  month     = Sep,
  number    = {6355},
  pages     = {995--1001},
  volume    = {357},
  doi       = {10.1126/science.aal3837},
  publisher = {American Association for the Advancement of Science (AAAS)},
}

@Article{Sch_fer_2020,
  author    = {Schäfer, Florian and Fukuhara, Takeshi and Sugawa, Seiji and Takasu, Yosuke and Takahashi, Yoshiro},
  journal   = {Nature Reviews Physics},
  title     = {Tools for quantum simulation with ultracold atoms in optical lattices},
  year      = {2020},
  issn      = {2522-5820},
  month     = Jul,
  number    = {8},
  pages     = {411--425},
  volume    = {2},
  doi       = {10.1038/s42254-020-0195-3},
  publisher = {Springer Science and Business Media LLC},
}

@Article{Altman_2021,
  author    = {Altman, Ehud and Brown, Kenneth R. and Carleo, Giuseppe and Carr, Lincoln D. and Demler, Eugene and Chin, Cheng and DeMarco, Brian and Economou, Sophia E. and Eriksson, Mark A. and Fu, Kai-Mei C. and Greiner, Markus and Hazzard, Kaden R.A. and Hulet, Randall G. and Kollár, Alicia J. and Lev, Benjamin L. and Lukin, Mikhail D. and Ma, Ruichao and Mi, Xiao and Misra, Shashank and Monroe, Christopher and Murch, Kater and Nazario, Zaira and Ni, Kang-Kuen and Potter, Andrew C. and Roushan, Pedram and Saffman, Mark and Schleier-Smith, Monika and Siddiqi, Irfan and Simmonds, Raymond and Singh, Meenakshi and Spielman, I.B. and Temme, Kristan and Weiss, David S. and Vučković, Jelena and Vuletić, Vladan and Ye, Jun and Zwierlein, Martin},
  journal   = {PRX Quantum},
  title     = {Quantum Simulators: Architectures and Opportunities},
  year      = {2021},
  issn      = {2691-3399},
  month     = Feb,
  number    = {1},
  pages     = {017003},
  volume    = {2},
  doi       = {10.1103/prxquantum.2.017003},
  publisher = {American Physical Society (APS)},
}

@Article{Daley_2022,
  author    = {Daley, Andrew J. and Bloch, Immanuel and Kokail, Christian and Flannigan, Stuart and Pearson, Natalie and Troyer, Matthias and Zoller, Peter},
  journal   = {Nature},
  title     = {Practical quantum advantage in quantum simulation},
  year      = {2022},
  issn      = {1476-4687},
  month     = Jul,
  number    = {7920},
  pages     = {667--676},
  volume    = {607},
  doi       = {10.1038/s41586-022-04940-6},
  publisher = {Springer Science and Business Media LLC},
}

@Article{Heyl_2013,
  author    = {Heyl, M. and Polkovnikov, A. and Kehrein, S.},
  journal   = {Physical Review Letters},
  title     = {Dynamical Quantum Phase Transitions in the Transverse-Field Ising Model},
  year      = {2013},
  issn      = {1079-7114},
  month     = Mar,
  number    = {13},
  pages     = {135704},
  volume    = {110},
  doi       = {10.1103/physrevlett.110.135704},
  publisher = {American Physical Society (APS)},
}

@Article{Jurcevic_2017,
  author    = {Jurcevic, P. and Shen, H. and Hauke, P. and Maier, C. and Brydges, T. and Hempel, C. and Lanyon, B. P. and Heyl, M. and Blatt, R. and Roos, C. F.},
  journal   = {Physical Review Letters},
  title     = {Direct Observation of Dynamical Quantum Phase Transitions in an Interacting Many-Body System},
  year      = {2017},
  issn      = {1079-7114},
  month     = Aug,
  number    = {8},
  pages     = {080501},
  volume    = {119},
  doi       = {10.1103/physrevlett.119.080501},
  publisher = {American Physical Society (APS)},
}

@Article{Fl_schner_2017,
  author    = {Fläschner, N. and Vogel, D. and Tarnowski, M. and Rem, B. S. and Lühmann, D.-S. and Heyl, M. and Budich, J. C. and Mathey, L. and Sengstock, K. and Weitenberg, C.},
  journal   = {Nature Physics},
  title     = {Observation of dynamical vortices after quenches in a system with topology},
  year      = {2017},
  issn      = {1745-2481},
  month     = Dec,
  number    = {3},
  pages     = {265--268},
  volume    = {14},
  doi       = {10.1038/s41567-017-0013-8},
  publisher = {Springer Science and Business Media LLC},
}

@Article{Calabrese_2005,
  author    = {Calabrese, Pasquale and Cardy, John},
  journal   = {Journal of Statistical Mechanics: Theory and Experiment},
  title     = {Evolution of entanglement entropy in one-dimensional systems},
  year      = {2005},
  issn      = {1742-5468},
  month     = Apr,
  number    = {04},
  pages     = {P04010},
  volume    = {2005},
  doi       = {10.1088/1742-5468/2005/04/p04010},
  publisher = {IOP Publishing},
}

@Article{Budich_2016,
  author    = {Budich, Jan Carl and Heyl, Markus},
  journal   = {Physical Review B},
  title     = {Dynamical topological order parameters far from equilibrium},
  year      = {2016},
  issn      = {2469-9969},
  month     = Feb,
  number    = {8},
  pages     = {085416},
  volume    = {93},
  doi       = {10.1103/physrevb.93.085416},
  publisher = {American Physical Society (APS)},
}

@Article{Andraschko_2014,
  author    = {Andraschko, F. and Sirker, J.},
  journal   = {Physical Review B},
  title     = {Dynamical quantum phase transitions and the Loschmidt echo: A transfer matrix approach},
  year      = {2014},
  issn      = {1550-235X},
  month     = Mar,
  number    = {12},
  pages     = {125120},
  volume    = {89},
  doi       = {10.1103/physrevb.89.125120},
  publisher = {American Physical Society (APS)},}

@book{Asboth2016,
author = {Asb{\'o}th, J{\'a}nos K. and Oroszl{\'a}ny, L{\'a}szl{\'o} and P{\'a}lyi, Andr{\'a}s},
title = {A Short Course on Topological Insulators},
publisher = {Springer},
year = {2016},
doi = {10.1007/978-3-319-25607-8}
}

\end{document}